\documentstyle[epsfig,12pt]{article}

\newcommand{\fig}[1]{Fig.\ref{#1}}
\newcommand{\tab}[1]{Table \ref{#1}}

\newcommand{\eqn}[1]{Eq.(\ref{#1})}

\newcommand{\ben}{\begin{enumerate}}
\newcommand{\een}{\end{enumerate}}
\newcommand{\bit}{\begin{itemize}}
\newcommand{\eit}{\end{itemize}}
\newcommand{\bc}{\begin{center}}
\newcommand{\ec}{\end{center}}
\newcommand{\bb}{\begin{bf}}
\newcommand{\eb}{\end{bf}}
\newcommand{\bsm}{\begin{small}}
\newcommand{\esm}{\end{small}}
\newcommand{\bns}{\begin{normalsize}}
\newcommand{\ens}{\end{normalsize}}
\newcommand{\bq}{\begin{equation}}
\newcommand{\eq}{\end{equation}}
\newcommand{\bqa}{\begin{eqnarray}}
\newcommand{\eqa}{\end{eqnarray}}

\newcommand{\vb}{\vspace*{2cm}}

\newcommand{\nn}{\nonumber}

\def\c2{\chi^2}
\def\SM{Standard Model\ }

\def\onehalf{{1\over 2}}

\def\demo{$\Delta\eta\mu \acute{o} \kappa \varrho \iota \tau o \varsigma$}

\def\erato{\verb+ERATO+}

\def\pr#1#2#3{ Phys. Rev. ${\bf{#1}}$ (#2) #3}
\def\prl#1#2#3{ Phys. Rev. Lett. ${\bf{#1}}$ (#2) #3}
\def\pl#1#2#3{ Phys. Lett. ${\bf{#1}}$ (#2) #3}
\def\prep#1#2#3{ Phys. Rep. ${\bf{#1}}$ (#2) #3}
\def\np#1#2#3{ Nucl. Phys. ${\bf{#1}}$ (#2) #3}
\def\zp#1#2#3{ Z. f. Phys. ${\bf{#1}}$ (#2) #3}
\def\ijmp#1#2#3{ Int. J. Mod. Phys. ${\bf{#1}}$ (#2) #3}

\def\ibid#1#2#3{ {\it ibid} ${\bf{#1}}$ (#2) #3}

\begin{document}
\input feynman
 
\pagestyle{empty}
 
\begin{flushright}
October 1998
\end{flushright}
 
\vspace*{2cm}
\bc\begin{LARGE}
{\bf
Single leptoquark production at high-energy $e^+ e^-$ colliders
}\\ \end{LARGE}
\vspace*{2cm}
{\Large
Costas G.~Papadopoulos
} \\[12pt]
Institute of Nuclear Physics, NCSR \demo, 15310 Athens, Greece
\\
\vb
ABSTRACT\\[12pt]  \ec
\begin{quote}

A study of the production and the decay of scalar and vector 
leptoquarks at high-energy $e^+e^-$ colliders is presented.
All tree-order contributions to single leptoquark 
production  have been calculated and incorporated in the Monte Carlo event
generator {\tt ERATO-LQ}.

\end{quote}
\vspace*{\fill}
\noindent\rule[0.in]{4.5in}{.01in} \\      
\vspace{.3cm}  
E-mail: {\tt Costas.Papadopoulos@cern.ch}.

\newpage
\pagestyle{plain}
\setcounter{page}{1}


\bc {\bf PROGRAM SUMMARY}\\[18pt]\ec
{\it Title of the program:} {\tt ERATO-LQ}. \\[12pt]
{\it Catalogue number:}  \\[12pt]
{\it Program obtainable from:} Dr. Costas G. Papadopoulos, 
Institute of Nuclear Physics,
NRCPS `Democritos', 15310 Athens, Greece.\\
Also at {\tt ftp://alice.nrcps.ariadne-t.gr/pub/papadopo/erato}.\\[12pt]
{\it Licensing provisions:} none\\[12pt]
{\it Computer for which the program is designed and others on 
which it has been tested:}
HP, IBM, ALPHA and SUN workstations.\\[12pt]
{\it Operating system under which the program has been tested:} UNIX\\[12pt]
{\it Programming language:} FORTRAN 77 and FORTRAN 90\\[12pt]
\\[12pt]
{\it Nature of physical problem:} 
Leptoquark states emerge in several theoretical frameworks, including
the Pati-Salam model, the MSSM, and composite models of quarks and
leptons. Leptoquark states coupled to the first generation
leptons and quarks can be studied in $e^+e^-$ colliders. 
The dominant mode, for leptoquark masses
close to the energy threshold, is the single leptoquark production.
Leptoquarks can be studied by selecting events with letpons and jets
and by investigating the invariant mass distribution of the lepton plus jet 
system.
\\[12pt]
{\it Method of solution:} 
Based on the four-fermion event generator {\tt ERATO}, we have included
all possible leptoquark contributions to the processes:
\[ 
e^-e^+\to \ell^-\ell^+ q \bar{q}
\]
and
\[
e^-e^+\to \ell^-\bar{\nu_\ell} q \bar{q}'\;.
\]
All tree-order Feynman graphs as well as all necessary phase space mappings 
have been calculated and incorporated in the present extension
{\tt ERATO-LQ}. Moreover interference terms with the tree-order SM 
contributions have been fully included. The decay of the leptoquarks 
has been treated exactly. An additional code providing the total cross
section in the resolved photon approximation has also been included.

\newpage

\par

\section{Introduction}

Search for new particles has been constantly the center of attention 
in contemporary high energy physics.
The initially reported excess of high $Q^2$ events 
at HERA by the H1~\cite{h1:1997}
and ZEUS~\cite{zeus:1997} collaborations~\footnote{For a recent 
review on leptoquark searches in high $Q^2$ events
at HERA see reference~\cite{newhera}} 
stimulated a substantial research
interest~\cite{italians,germans,others,hewrizzo} on the possible 
existence of new particles of leptoquark type.
Such particles emerge in several contexts, including the Pati-Salam 
model~\cite{pati-salam} and the grand-unified supersymmetric 
theories~\cite{gunif}.
Leptoquark-type interactions, described by higher-dimensional contact terms,
are naturally incorporated within the idea of compositeness of quarks
and leptons~\cite{compos}.

It is the aim of this paper to present a complete description 
of the relevant four fermion final states, associated with 
leptoquark production at high-energy $e^+e^-$ colliders. 
This automatically includes pair leptoquark production as well
as single leptoquark production channels. 
As a first application we present results concerning 
single leptoquark production at LEP2.    
 
We start the presentation with 
the most general effective Lagrangian~\cite{brw}, 
which is $SU(3)\times SU(2)\times U(1)$
invariant, baryon- and lepton-number conserving as well as family diagonal,
in order to avoid the most severe constraints from low-energy 
considerations~\cite{lowenergy}.
It can be written as 
\bqa 
{\cal L}_{F=-2}&=&
\left(g_{0L}\;\bar{q}^c_L i\tau_2 \ell_L+g_{0R} \;\bar{u}^c_R e_R\right)
S_0
+
 \tilde{g}_{0R}\;\bar{d}^c_R e_R \tilde{S}_0
+g_{1L}\;\bar{q}^c_L i\tau_2 \mbox{\boldmath $\vec{\tau}$} \ell_L \vec{S}_1
\nn \\
&+& 
\left( g_{\onehalf L}\;\bar{d}^c_L\gamma^\mu\ell_L + 
g_{\onehalf R}\;\bar{q}^c_L \gamma^\mu e_R \right)
V_{\onehalf\mu}
+
\tilde{g}_{\onehalf L}\;\bar{u}^c_R \gamma^\mu \ell_L \tilde{V}_{\onehalf\mu} 
+ \mbox{cc}
\label{f=2}
\eqa
\bqa 
{\cal L}_{F=0}&=&
\left(h_{\onehalf L}\;\bar{u}_R \ell_L
+h_{\onehalf R} \;\bar{q}_L i\tau_2 e_R\right)
S_\onehalf
+\tilde{h}_{\onehalf L}\;\bar{d}_R \ell_L\tilde{S}_\onehalf
+\tilde{h}_{0R}\;\bar{u}_R \gamma^\mu e_R \tilde{V}_{0\mu}
\nn \\
&+&
\left( h_{0L}\;\bar{q}_L \gamma^\mu \ell_L + 
        h_{0R}\;\bar{d}_R \gamma^\mu e_R \right)V_{0\mu}
+h_{1L}\bar{q}_L \mbox{\boldmath $\vec{\tau}$}\gamma^\mu \ell_L
\vec{V}_{1\mu} 
+ \mbox{cc}
\label{f=0}
\eqa
where $F=3B+L$ denotes the fermion number. The gauge-interaction 
terms are given by 
\bq
{\cal L}= \left(D_\mu\mbox{\boldmath $\Phi$}\right)^\dagger
 D^\mu\mbox{\boldmath $\Phi$}-M^2 \mbox{\boldmath $\Phi$}^\dagger
 \mbox{\boldmath $\Phi$}
\eq 
for scalars and 
\bq
{\cal L}= -\onehalf G^\dagger_{\mu\nu} G^{\mu\nu}
+M^2 \mbox{\boldmath $\Phi$}_\mu^\dagger\mbox{\boldmath $\Phi$}^\mu
\eq 
for vectors, with 
\bq
G_{\mu\nu}=D_\mu \mbox{\boldmath $\Phi$}_\nu -D_\nu \mbox{\boldmath $\Phi$}_\mu 
\eq
and the $SU(2)_L\times U(1)_Y$ covariant derivative given by
\bq
D_\mu = \left[\partial_\mu -ie Q^\gamma A_\mu -ie Q^Z Z_\mu-ie Q^W 
(W^+_\mu I^+ +W^-_\mu I^-)\right]
\eq 
where $I^\pm$ are the $SU(2)$ generators in the representation of the
corresponding leptoquark state.
The electroweak charges are given by
\bq 
Q^Z=\frac{I_3-Q^\gamma \sin^2 \theta_w}{\cos\theta_w\;\sin\theta_w}
\eq
and 
\bq
Q^W=\frac{1}{\sqrt{2}\sin\theta_w}\;\;.
\eq
In order to account for a more general structure in the vector
leptoquark case one can also add the so called `anomalous couplings' terms
as for instance the `magnetic' dipole one given by
\bq
{\cal L}_{extra} = -i  \sum_{V=\gamma,Z} g_V \kappa_V \mbox{\boldmath $\Phi$}^\dagger_\mu
V^{\mu\nu}\mbox{\boldmath $\Phi$}_\nu
\eq
parametrized in terms of $\kappa_\gamma$ and $\kappa_Z$.


\section{Single leptoquark production}

Leptoquark contributions to $e^+e^-$ processes can be classified into 
real and virtual ones. To the lowest order, virtual
contributions proceed via the reaction 
$e^+e^-\to q\bar{q}$~\cite{italians,germans,hewrizzo}, 
whereas for real ones we have the single leptoquark
production mode~\cite{godfrey,french,turkey} and for light enough masses, 
or high enough energies, the pair leptoquark production~\cite{bluml+ruckl}.


\begin{figure}[th]
\bc
\unitlength=0.01pt
\bigphotons
\begin{picture}(40000,25000)
\global\Xone=5000
\global\Xtwo=2500
\global\Yone=3
\global\Ytwo=4
\global\Xthree=0
\global\Ythree=0

\pfrontx = 1000
\pfronty = 22000

\drawline\fermion[\E\REG](\pfrontx,\pfronty)[5000]
\global\advance \pfrontx by -800
\put(\pfrontx,\pfronty){$e$}
\Xone=\pmidx
\Yone=\pmidy
\drawline\fermion[\E\REG](\pbackx,\pbacky)[5000]
\global\advance \pbackx by  800
\put(\pbackx,\pbacky){$q$}

\Xtwo=\pfrontx
\Ytwo=\pfronty

\drawline\photon[\NE\REG](\Xone,\Yone)[4]
\drawline\fermion[\NE\REG](\pbackx,\pbacky)[2000]
\global\advance \pbackx by  800
\put(\pbackx,\pbacky){$e(p_3)$}
\drawline\fermion[\SE\REG](\pfrontx,\pfronty)[2000]
\global\advance \pbackx by  800
\put(\pbackx,\pbacky){$e$}

\drawline\scalar[\S\REG](\Xtwo,\Ytwo)[4]
\global\advance \pmidx by -2500
\global\advance \pmidy by -500
\put(\pmidx,\pmidy){$LQ$}

\drawline\fermion[\E\REG](\pbackx,\pbacky)[5000]
\global\advance \pbackx by 500
\put(\pbackx,\pbacky){$q$}

\drawline\fermion[\W\REG](\pfrontx,\pfronty)[5000]
\global\advance \pbackx by -2500
\put(\pbackx,\pbacky){$e(p_2)$}

\pfrontx = 13000
\pfronty = 22000

\drawline\fermion[\E\REG](\pfrontx,\pfronty)[5000]
\drawline\fermion[\E\REG](\pbackx,\pbacky)[5000]

\Xtwo=\pfrontx
\Ytwo=\pfronty

\drawline\photon[\NE\REG](\Xone,\Yone)[4]
\drawline\fermion[\NE\REG](\pbackx,\pbacky)[2000]
\drawline\fermion[\SE\REG](\pfrontx,\pfronty)[2000]

\drawline\scalar[\S\REG](\Xtwo,\Ytwo)[4]

\drawline\fermion[\E\REG](\pbackx,\pbacky)[5000]

\drawline\fermion[\W\REG](\pfrontx,\pfronty)[5000]
\Xone=\pmidx
\Yone=\pmidy
\drawline\photon[\SE\REG](\Xone,\Yone)[4]
\drawline\fermion[\NE\REG](\pbackx,\pbacky)[2000]
\drawline\fermion[\SE\REG](\pfrontx,\pfronty)[2000]

\pfrontx = 26000
\pfronty = 22000

\drawline\fermion[\E\REG](\pfrontx,\pfronty)[5000]
\drawline\fermion[\E\REG](\pbackx,\pbacky)[5000]
\Xone=\pmidx
\Yone=\pmidy

\Xtwo=\pfrontx
\Ytwo=\pfronty

\drawline\photon[\NE\REG](\Xone,\Yone)[4]
\drawline\fermion[\NE\REG](\pbackx,\pbacky)[2000]
\drawline\fermion[\SE\REG](\pfrontx,\pfronty)[2000]

\drawline\scalar[\S\REG](\Xtwo,\Ytwo)[4]

\drawline\fermion[\E\REG](\pbackx,\pbacky)[5000]

\drawline\fermion[\W\REG](\pfrontx,\pfronty)[5000]

\pfrontx = 1000
\pfronty = 8000

\drawline\fermion[\E\REG](\pfrontx,\pfronty)[5000]
\drawline\fermion[\E\REG](\pbackx,\pbacky)[5000]

\Xtwo=\pfrontx
\Ytwo=\pfronty

\drawline\scalar[\S\REG](\Xtwo,\Ytwo)[4]

\drawline\fermion[\E\REG](\pbackx,\pbacky)[5000]
\Xone=\pmidx
\Yone=\pmidy

\drawline\fermion[\W\REG](\pfrontx,\pfronty)[5000]

\drawline\photon[\NE\REG](\Xone,\Yone)[4]
\drawline\fermion[\NE\REG](\pbackx,\pbacky)[2000]
\drawline\fermion[\SE\REG](\pfrontx,\pfronty)[2000]

\pfrontx = 13000
\pfronty = 8000

\drawline\fermion[\E\REG](\pfrontx,\pfronty)[5000]
\drawline\fermion[\E\REG](\pbackx,\pbacky)[5000]

\Xtwo=\pfrontx
\Ytwo=\pfronty

\seglength 800
\drawline\scalar[\S\REG](\Xtwo,\Ytwo)[3]
\seglength 800
\drawline\scalar[\S\REG](\pbackx,\pbacky)[3]

\Xtwo=\pbackx
\Ytwo=\pbacky
\drawline\photon[\E\REG](\pfrontx,\pfronty)[3]
\drawline\fermion[\NE\REG](\pbackx,\pbacky)[2000]
\drawline\fermion[\SE\REG](\pfrontx,\pfronty)[2000]

\drawline\fermion[\E\REG](\Xtwo,\Ytwo)[5000]
\drawline\fermion[\W\REG](\pfrontx,\pfronty)[5000]

\pfrontx = 26000
\pfronty = 8000

\drawline\fermion[\E\REG](\pfrontx,\pfronty)[5000]
\drawline\fermion[\E\REG](\pbackx,\pbacky)[5000]

\Xtwo=\pfrontx
\Ytwo=\pfronty

\seglength=1100
\drawline\scalar[\S\REG](\Xtwo,\Ytwo)[2]
\drawline\fermion[\E\REG](\pbackx,\pbacky)[4000]
\drawline\fermion[\S\REG](\pfrontx,\pfronty)[2000]
\drawline\fermion[\E\REG](\pbackx,\pbacky)[4000]
\seglength=1100
\drawline\scalar[\S\REG](\pfrontx,\pfronty)[2]

\drawline\fermion[\E\REG](\pbackx,\pbacky)[5000]
\drawline\fermion[\W\REG](\pfrontx,\pfronty)[5000]

\end{picture}
\ec
\caption[.]{\verb+EQ20+, \verb+EQ10+ and \verb+EQ5+ classes. 
The curly lines can be $\gamma$, $Z$ or
$W$, with the appropriate changes in the fermion species. 
The other half of the graphs are obtained, as usually, by the interchange 
$e(p_2)\leftrightarrow e(p_3)$.}
\label{lqgen}
\end{figure}
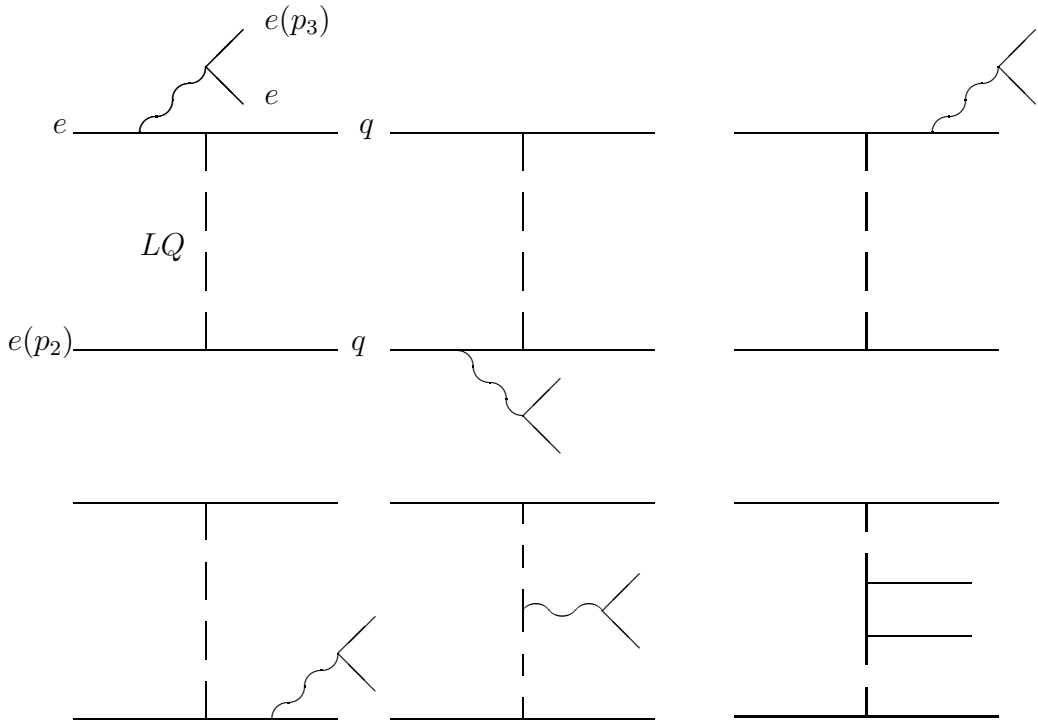

Leptoquark production at high-energy $e^+e^-$ collisions
results to a four-fermion final state. 
We have calculated all tree-order contributions, 
including all interference terms with the Standard Model,
to the following `semileptonic' reactions
\bq 
e^-e^+\to \ell^-\ell^+ q \bar{q}
\label{ncfs}
\eq
and
\bq
e^-e^+\to \ell^-\bar{\nu_\ell} q \bar{q}'\;,
\label{ccfs}
\eq
where $\ell=e,\mu,\tau$. A full description of the MC generator
\erato\  can be found in 
references~\cite{erato,lep2gen}. A brief description of its present
extension is given in the Appendix. 
Leptoquark contributions to $\nu_\ell \bar{\nu}_\ell q \bar{q}$ and
$q \bar{q} q' \bar{q}'$ final states have not been considered here,
since, as we will see, they are relatively suppressed.

For the first reaction, \eqn{ncfs}, with $\ell=e$ ($\ell=\mu,\tau$), the \SM Feynman 
graphs are of the 
\verb+NC48+ (\verb+NC24+) type\footnote{For the nomenclature about the four-fermion 
production diagrams see reference~\cite{lep2gen}.},
whereas the leptoquark contribution
can be classified in two categories: the \verb+LQ22+ (\verb+LQ11+), which is
diagrammatically identical to the \SM \verb+CC22+ (\verb+CC11+) with
the $W$ boson replaced by a leptoquark, and a new category, called
\verb+EQ20+, which contributes only when $\ell=e$. The generic graphs 
belonging to the latter class are shown
in \fig{lqgen}. 
Moreover there are nine scalar and nine vector leptoquark states
contributing to these processes, namely 
\[
S^L_{0},S^R_{0},\tilde{S}^R_{0},S^L_{1+},S^L_{10},S^L_{\onehalf +}
,S^R_{\onehalf +},S^R_{\onehalf -}\;\;\mbox{and}\;\;\tilde{S}^R_{\onehalf +}\;
\] 
and 
\[
V^L_{\onehalf +}
,V^R_{\onehalf +},V^R_{\onehalf -},\tilde{V}^R_{\onehalf +}
V^L_{0},V^R_{0},\tilde{V}^R_{0},V^L_{1+}\;\;\mbox{and}\;\;V^L_{10}\;,
\] 
with the obvious notation $X^a_{bc}$, where
$a=L,R$ is the helicity of the lepton, $b=I$ is the isospin of the
leptoquark and $c=I_3$ stands for its third component.

For the reaction \eqn{ccfs}, \SM contributions belong to the well known 
\verb+CC20+(\verb+CC10+) category, whereas those of leptoquarks fall into three
different classes. The first one is the \verb+LQ20+(\verb+LQ10+), 
in close analogy to the previous case. The other two,
the \verb+EQ10+  and \verb+EQ6+, 
\fig{lqgen}, receive contributions only from the first generation leptoquarks.
In the \verb+LQ20+(\verb+LQ10+) class only $S^L_{0},\vec{S}^L_1$ and 
$U^L_{0},\vec{U}^L_1$ states contribute,
whereas in the classes \verb+EQ10+ and \verb+EQ6+ we have contributions from
$S^L_{1}$, $V^L_{1}$
and $S^R_{\onehalf}$, $V^R_{\onehalf}$ respectively.

In addition to the calculation of all tree-order Feynman graphs, 
we have also employed all phase-space mappings~\cite{excalibur,erato}, which 
are necessary to cover all the kinematical regions where leptoquarks have 
the most substantial contributions. Initial state radiation (ISR)
has also been included in the structure function
approach~\cite{lep2ww}.
Our calculation provides therefore the necessary framework to investigate
real leptoquark production at high-energy $e^+e^-$ colliders, 
including both pair- and single-production modes, for all
leptoquark types, as given by \eqn{f=2}-\eqn{f=0}, and for all generations.
Nevertheless we find it convenient, to focus our present analysis on 
the first generation
leptoquarks. It should be mentioned, however, that all necessary
contributions for the study of the 2nd and
3rd generation leptoquarks at LEP2
have been fully accounted for in the present calculation.

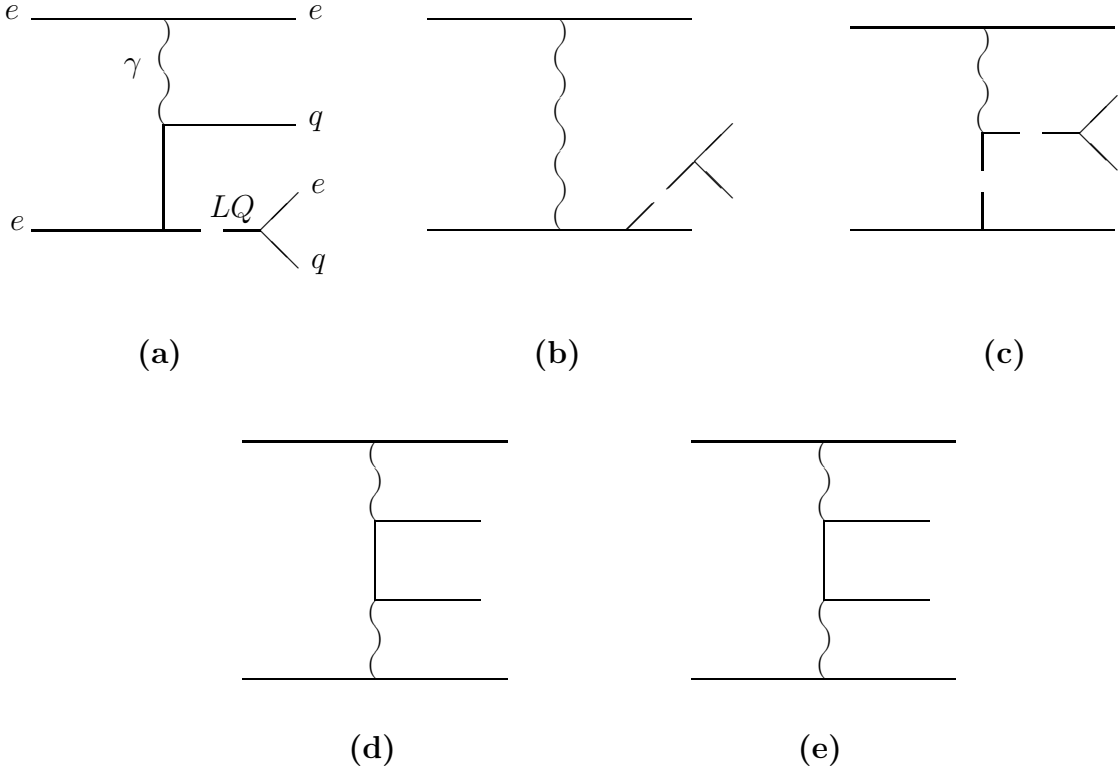
\begin{figure}[th]
\bc
\unitlength=0.01pt
\begin{picture}(40000,30000)
\global\Xone=5000
\global\Xtwo=2500
\global\Yone=3
\global\Ytwo=4
\global\Xthree=0
\global\Ythree=0

\Xthree = 0
\Ythree = 20000
\pfrontx = \Xthree
\pfronty = \Ythree

\drawline\fermion[\E\REG](\pfrontx,\pfronty)[5000]
\global\advance \pfrontx by -800
\put(\pfrontx,\pfronty){$e$}

\drawline\scalar[\E\REG](\pbackx,\pbacky)[2]
\global\advance \pmidy by  500
\global\advance \pmidx by -100
\put(\pmidx,\pmidy){$LQ$}

\Xone=\pfrontx
\Yone=\pfronty
\drawline\fermion[\NE\REG](\pbackx,\pbacky)[2000]
\global\advance \pbackx by 500
\put(\pbackx,\pbacky){$e$}
\drawline\fermion[\SE\REG](\pfrontx,\pfronty)[2000]
\global\advance \pbackx by 500
\put(\pbackx,\pbacky){$q$}
\drawline\fermion[\N\REG](\Xone,\Yone)[4000]
\drawline\fermion[\E\REG](\pbackx,\pbacky)[5000]
\global\advance \pbackx by 500
\put(\pbackx,\pbacky){$q$}
\drawline\photon[\N\REG](\pfrontx,\pfronty)[\Ytwo]
\global\advance \pmidx by -1500
\put(\pmidx,\pmidy){$\gamma$}
\drawline\fermion[\E\REG](\photonbackx,\photonbacky)[5000]
\global\advance \pbackx by 500
\put(\pbackx,\pbacky){$e$}
\drawline\fermion[\W\REG](\photonbackx,\photonbacky)[5000]
\global\advance \pbackx by -1000
\put(\pbackx,\pbacky){$e$}

\global\advance \pfrontx by 10000

\drawline\fermion[\E\REG](\pfrontx,\pfronty)[5000]
\drawline\fermion[\E\REG](\pbackx,\pbacky)[5000]
\drawline\photon[\S\REG](\pfrontx,\pfronty)[8]
\drawline\fermion[\W\REG](\photonbackx,\photonbacky)[5000]
\drawline\fermion[\E\REG](\pfrontx,\pfronty)[5000]
\drawline\fermion[\NE\REG](\pmidx,\pmidy)[1500]
\global\advance \pbackx by 500
\global\advance \pbacky by 500
\drawline\fermion[\NE\REG](\pbackx,\pbacky)[1500]

\drawline\fermion[\NE\REG](\pbackx,\pbacky)[2000]
\drawline\fermion[\SE\REG](\pfrontx,\pfronty)[2000]

\global\advance \photonbackx by 16000 

\drawline\fermion[\W\REG](\photonbackx,\photonbacky)[5000]
\drawline\fermion[\E\REG](\photonbackx,\photonbacky)[5000]
\drawline\scalar[\N\REG](\photonbackx,\photonbacky)[2]
\drawline\scalar[\E\REG](\pbackx,\pbacky)[2]
\Xone=\pfrontx
\Yone=\pfronty
\drawline\fermion[\NE\REG](\pbackx,\pbacky)[2000]
\drawline\fermion[\SE\REG](\pfrontx,\pfronty)[2000]
\drawline\photon[\N\REG](\Xone,\Yone)[4]
\drawline\fermion[\W\REG](\photonbackx,\photonbacky)[5000]
\drawline\fermion[\E\REG](\photonbackx,\photonbacky)[5000]

\pfrontx = 8000
\pfronty = 3000

\drawline\fermion[\E\REG](\pfrontx,\pfronty)[5000]
\drawline\fermion[\E\REG](\pbackx,\pbacky)[5000]
\drawline\photon[\N\REG](\pfrontx,\pfronty)[3]
\drawline\fermion[\E\REG](\pbackx,\pbacky)[4000]
\drawline\fermion[\N\REG](\pfrontx,\pfronty)[3000]
\drawline\fermion[\E\REG](\pbackx,\pbacky)[4000]
\drawline\photon[\N\REG](\pfrontx,\pfronty)[3]
\drawline\fermion[\E\REG](\pbackx,\pbacky)[5000]
\drawline\fermion[\W\REG](\pfrontx,\pfronty)[5000]

\global\advance \pfrontx by 12000 
\pfronty = 3000

\drawline\fermion[\E\REG](\pfrontx,\pfronty)[5000]
\drawline\fermion[\E\REG](\pbackx,\pbacky)[5000]
\drawline\photon[\N\REG](\pfrontx,\pfronty)[3]
\drawline\fermion[\E\REG](\pbackx,\pbacky)[4000]
\drawline\fermion[\N\REG](\pfrontx,\pfronty)[3000]
\drawline\fermion[\E\REG](\pbackx,\pbacky)[4000]
\drawline\photon[\N\REG](\pfrontx,\pfronty)[3]
\drawline\fermion[\E\REG](\pbackx,\pbacky)[5000]
\drawline\fermion[\W\REG](\pfrontx,\pfronty)[5000]

\put(4000,15000){\bf (a)}
\put(19000,15000){\bf (b)}
\put(36000,15000){\bf (c)}
\put(12000,0){\bf (d)}
\put(29000,0){\bf (e)}

\end{picture}

\ec
\caption[.]{The subset of Feynman graphs giving the dominant contribution 
to single leptoquark production: (a,b,c) signal, (e,d) background.}
\label{slep}
\end{figure}

Although leptoquark contributions appear in a large number of graphs
only a small part of them give a sizable contribution to the production
cross-section, depending of course on the leptoquark parameters such as
the mass and the Yukawa coupling. 
The first important consideration is coming from 
the total width of the leptoquark, which is given by
\bq
\Gamma_J=f_J\;\frac{M}{8\pi}\;\sum_{i=1}^{N_{ch}}\lambda_i^2 
\eq
where $J=0,1$ is the spin of the particle, 
$f_0=\frac{1}{2}$ and $f_1=\frac{1}{3}$ 
and $N_{ch}=1\;\mbox{or}\;2$ depending on how many channels
are available. As is evident for a leptoquark mass of the ${\cal O}(200$~GeV) 
the typical width is of the order of a few hundreds of MeV. This is
mainly due to the fact that leptoquarks couple to ordinary fermions
in a very restricted way, which is not generally true for other
types of leptoquark couplings, like those appearing in R-parity violating
MSSM~\cite{italians}.
As a consequence of the narrowness of these states 
the interference terms among resonant and non-resonant contributions 
are rather suppressed.

The second remark is that, for leptoquark masses $M\ge \sqrt{s}/2$, the
single leptoquark production mode becomes dominant as far as the
first generation leptoquarks are concerned. This proceeds mainly via
the $t$-channel graphs, shown in \fig{slep}, and their contribution 
comes mainly from the collinear electrons (positrons). This can be described 
either by the Weizs\"{a}cker-Williams~\cite{ww-app} approximation 
or by integrating over
the momentum of the final state electron using the leading logarithmic 
approximation as described in reference~\cite{excalibur}.
We have implemented {\it both} approaches, and checked that the results agree
within less that 10\%, which is rather typical for the leading logarithmic (LL)
approximation we are essentially employing in both cases. As far as the 
Weizs\"{a}cker-Williams
spectrum is concerned, we have used the following form:
\bq
f_{\gamma/e}(x,s)=\frac{\alpha}{2\pi}\left[
\frac{1+(1-x)^2}{x} \ln\left(\frac{s}{m_e^2}\frac{(1-x)^2}{(2-x)^2}\right)
+x\ln\frac{2-x}{x}+\frac{2(x-1)}{x}\right]\;\;.
\eq
It should be mentioned that this type of contributions, which are 
substantially enhanced due to the $t$-channel photon exchange, are
only relevant for the first generation leptoquarks and for the
semileptonic reactions, \eqn{ncfs}-\eqn{ccfs}, under consideration. 
On the other hand, first generation leptoquark contributions to 
$\nu_\ell \bar{\nu}_\ell q \bar{q}$ and
$q \bar{q} q' \bar{q}'$ final states are relatively suppressed, due to the
absence of the $t$-channel photon exchange enhancement.

\begin{figure}[th]
\begin{center}
\mbox{
\epsfig{file=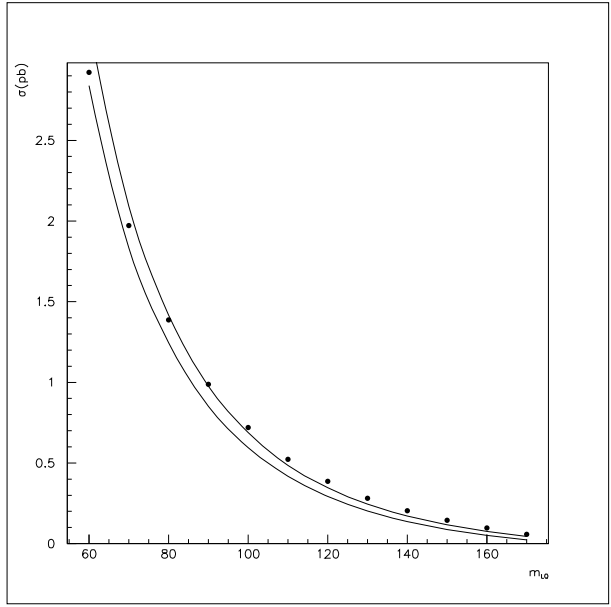,height=8cm,width=8cm}}
\caption[.]{Total cross section for single-leptoquark production, $S_0$ (u-type)
as a function of the leptoquark mass, for $\lambda/e=\frac{1}{2}$. 
The solid lines represent the resolved photon contribution (upper: SAS-G 1D,
lower: DO-G Set1) whereas points refer to the perturbative result with 
light quark mass 300 MeV.}
\label{grv-pt}
\end{center}
\end{figure}

Finally, as the first graph of \fig{slep} has a mass singularity
in the quark propagator, special care has to be paid in the integration
in this specific case. As before the LL-approximation scheme has been
employed. It is worth to mention that comparing with the exact
result for the total cross-section~\cite{french}, taking into account 
the quark masses\footnote{For the light quarks we take $m_{(u,d)}=300$MeV.}, 
and properly 
integrating over the photon spectrum, both results agree to less 
than a few percent. Moreover in order to have a quantitative estimate 
of the QCD effects, due to the presence of the light quark mass singularity,
we compare the perturbative results with the resolved photon contribution.
In the latter case the total cross section for the single-leptoquark
production is given by 
\bq
\sigma(e^++e^-\to S_0+X) = 
\int dx\; dz f_{\gamma/e}(x,s) 
f_{q/\gamma}( z,M_{LQ}^2 ) \hat{\sigma}(xzs)
\eq
which can be written as
\bq
\sigma(e^++e^-\to S_0+X) = \frac{\pi^2 \alpha_{em}}{s}
\left(\frac{\lambda}{e}\right)^2
\int_{M_{LQ}^2/s}^{1} \frac{dx}{x} f_{\gamma/e}(x,s) 
f_{q/\gamma}( M_{LQ}^2/(xs),M_{LQ}^2 )
\label{respho}
\eq
where $f_{q/\gamma}$ is the structure function of the quark inside the
photon. In \fig{grv-pt} we show the cross section as a function
of the leptoquark mass for $\sqrt{s}=192$~GeV, for two different choices of
the structure function parametrizations~\cite{PDF} compared 
to the perturbative calculation. 
We see that, even in the 
total cross section, one can safely trust the perturbative estimate. 
Moreover this exercise shows us that an error of the order of 10\% to 20\% 
is to be expected, whereas the differences between 
the two schemes are concentrated in the mass region near the kinematical
threshold.


\section{Results \& Discussion}

\begin{figure}[th]
\begin{center}
\mbox{
\epsfig{file=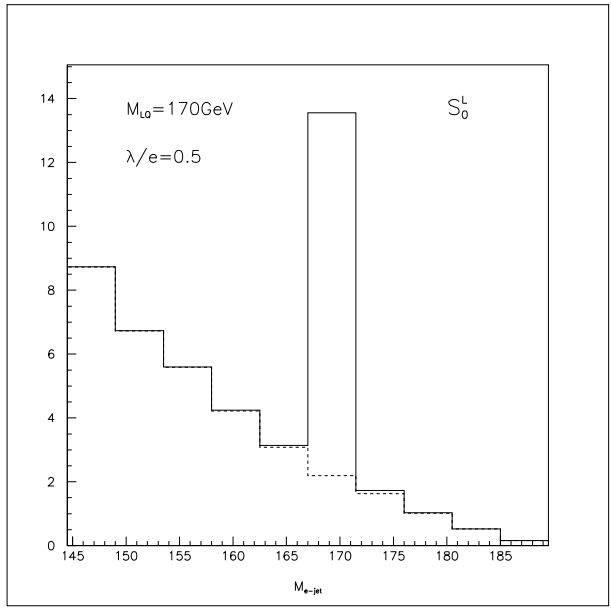,height=8cm,width=8cm}}
\caption[.]{The $m_{e-jet}$ invariant mass distribution 
in number of events per bin.}
\label{nclq}
\end{center}
\end{figure}

After completing the description of our calculation, we now come to 
the physics picture emerging from it.
In the case of \verb+NC+-type final states, \eqn{ncfs}, the signal 
has the typical
structure of a lepton and a jet balancing each other's transverse momentum
with some hadronic activity in the forward~(backward) region. In the irreducible
\SM background, on the other hand, the angular distribution of the 
positron~(electron) is peaking in the backward~(forward) direction. 
In order to suppress
as much as possible the background, without lowering signal's
contribution, we have employed the following set of cuts:
\bq
m_{e-jet}\ge 140\;\mbox{GeV}\;,\;\;
5^o\le\mbox{max}(\theta_{j1},\theta_{j2})\le 175^o\;,\;\;
20^o\le\theta_e\le 160^o\;,\;\;
E_e\ge 5\;\mbox{GeV}\;.
\eq
Among them the most important ones are the cut on the angle of
the observed positron (electron) and the cut on the electron-jet invariant
mass, $m_{e-jet}$.
In \fig{nclq} we show the invariant mass, $m_{e-jet}$, distribution
for leptoquark mass $M=170$~GeV and $\lambda/e=0.5$, with an integrated
luminosity $L=500$~pb$^{-1}$.   

\begin{figure}[th]
\begin{center}
\mbox{
\epsfig{file=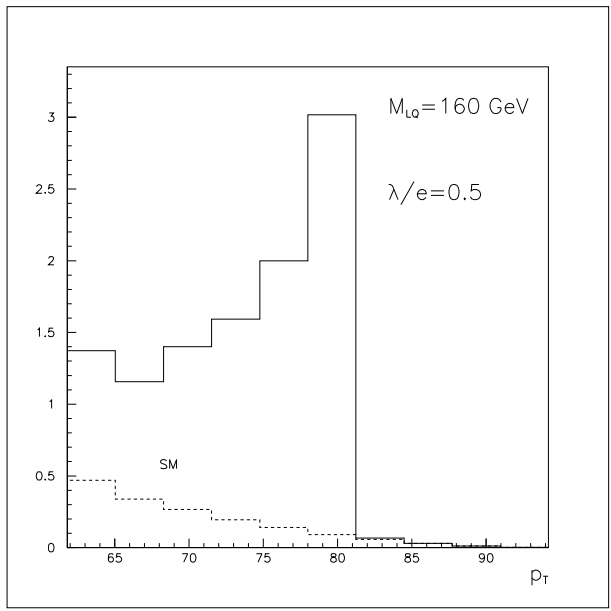,height=8cm,width=8cm}}
\caption[.]{The $p_T$ distribution in number of events per bin.}
\label{cclq}
\end{center}
\end{figure}

In the case of
\verb+CC+-type final states, \eqn{ccfs}, the signal, which is much more spectacular, 
consists only 
of one very energetic jet and a large amount of missing transverse energy.
The only cuts employed in this case are
\bq 
\mbox{max}(p_{T1},p_{T2})\ge 60\; \mbox{GeV}\;\;,\;\;
\mbox{min}(\theta_{j1},\theta_{j2})\le 5^o \;\mbox{or}\;\ge 175^o\;,
\eq
where $p_T$ is the transverse energy of the jet. The second cut is
employed in order to reduce the main irreducible background contribution, 
coming from single $W$ production, in which case two energetic jets,
with an invariant mass ${\cal O}( M_W)$, would be present in the final state
at relatively large angles with respect to the beam. 
Moreover, as the only kinematical information available in this channel 
is the jet momentum,
one cannot fully reconstruct the
mass of the leptoquark state, due to  the missing neutrino energy\footnote{However
the leptoquark mass can still be determined by the endpoint of the $p_T$ spectrum.}. 
In \fig{cclq} we show
the $p_T$ spectrum of the signal as well as that of the irreducible \SM
background.  
  
In summary, we have presented a calculation of leptoquark contributions
to two- and four-fermion final states at $e^+e^-$ collisions, which has
been incorporated in the MC event generator \erato. Taking into account the
\SM irreducible backgrounds we have presented a first study 
of single leptoquark production, in both its
\verb+NC+ or \verb+CC+ channel at LEP2 energies.  
\\[64pt]


\noindent{\Large\bf Appendix: The program {\tt ERATO-LQ}}
\\[24pt]

In this appendix we briefly describe the program {\tt ERATO-LQ}.
The skeleton of the program follows closely the skeleton of the
mother program {\tt ERATO}~\cite{erato}.
The distributed code contains two main directories, {\tt ffiles} where
all {\tt FORTRAN} files are stored and {\tt run} where the input files
as well as the {\tt makefile} resides.
There are two main {\tt FORTRAN} files, {\tt lqq.f} and {\tt vud.f}
corresponding to $e^+e^-\to e^- e^+ q\bar{q}$ and 
$e^+e^-\to e^- \bar{\nu_e} u \bar{d}$ processes respectively.
The files {\tt algor.f}, {\tt alice,f} and {\tt common.f} contains
all necessary routines  to make the program run
\footnote{Random number generator and the gamma function $\Gamma(x)$
are obtained from the Cern Library, CERNLIB.}.
The input files look like:

\begin{verbatim}
7,1,1,0         !process,scalar or vector
3               !iterations
1               !ISR
128.07 0.2310309 91.1888 2.4974 80.23 2.033  !input parameters
1               !itotal
-1. 0.1         !cmax(cmin) cmas
192             !energy
50000           !nev
145. 0.5        !LQ parameters : mass and l/e
10000.          !scut = (mass-10)**2
2000            !IEV
1               !iopt :=0 no optimization
4000 6000 10000 15000 20000 30000 120000 250000 500001
100.            !cut on invariant mass l+q= mass-10
0.3             !mass of light quark
20. 5. 5. 5. 5. 25.    !angles=degrees: clmax cjmax eelmin ejmin c4max cmas
1               !naive
\end{verbatim}
which is very much the same as the one used by {\tt ERATO}~\cite{erato}.
The new variables used in the present code, in addition to those
already described in reference~\cite{erato},  are:
\begin{itemize}
\item {\tt IPRO} 
\\ This is used to select the corresponding process: {\tt IPRO=5} selects
 $e^+e^-\to e^- e^+ u\bar{u}$, {\tt IPRO=6} $e^+e^-\to e^- e^+ d\bar{d}$.
\item{\tt ICHOICE} 
\\ This is used to select the specific leptoquark (charge, decay mode)
as described in \tab{choicess} and \tab{choicesv}.
\item{\tt ISCALAR}
\\{\tt ISCALAR=1} if a scalar leptoquark is selected.
\item{\tt IVECTOR}
\\{\tt IVECTOR=1} if a vector leptoquark is selected.
\item{\tt RLQMAS}
\\ The mass of the leptoquark.
\item{\tt CLQ}
\\ The leptoquark coupling, $g_L$($g_R$), as described in 
\tab{choicess} and \tab{choicesv},
normalized to the electromagnetic one, i.e, $g_L\leftarrow 1$ means $g_L=e$, 
$e=\sqrt{4\pi \alpha_{em}}$.
\item{\tt QMAS}
\\ This is the value of the light quark mass used as regulator of the
IR divergences of the amplitude. By default is set equal to 300~MeV.
\end{itemize}  

The final-state kinematics is represented by the three momenta,
$p_4(e^+)$, $p_3(u)$ and $p_{10}(d)$. 
Variable {\tt clmax} is the minimum angle allowed between the
momentum $p_4$ and the beam; {\tt cjmax} the minimum angle (in degrees) 
allowed between 
the hardest jet ($p_3$ or $p_{10}$) and the beam; {\tt eelmin} is the
minimum energy of $p_4$; {\tt ejmin} is the minimum energy of the
hardest jet; {\tt c4max} is the minimum angular separation between
the hardest jet and the lepton; {\tt cmas} is the minimum invariant
mass of the two-jet system. If {\tt NITER=3} then the output provides
cross sections for the signal process ({\tt ITER=1}) and for the
Standard Model contribution ({\tt ITER=2}). In addition, the total cross
section for on-shell leptoquark production is given ({\tt ITER=3}), where
the decay process of the leptoquark has not been taken into account and
no phase-space cut has been applied.

Finally a typical output of the {\tt lqq.f} code, corresponding to the input
described above, looks as follows:

\begin{verbatim}
                TOTAL SIGMA IN NB       ERROR
 LQ ERATO      1.749165382052100E-05 3.148577170490561E-07
 LQ:  on-shell 3.939744840638816E-05 3.045651348671309E-07
 SM            1.206926939341498E-04 7.367659234150230E-05
   [TOTAL,PASS,FAIL,ICCUT]   :   50000  9951 40049 29398  1152  2645  6854

\end{verbatim} 

In addition a programme called {\tt phot.f} which calculates the total
cross section from the resolved photon contribution, \eqn{respho},
is provided. All the leptoquark coupling (see \tab{choicess} and \tab{choicesv})
have been incorporated. A link to the standard parton distribution
functions library, {\tt PDFLIB}, is assumed.
\begin{table}[ht]
\bc
\begin{tabular}{|c|c|c|c|c|c|}
\hline &&&&&\\[-5pt]
LQ & {\tt ICHOICE} & Q & Decay Mode & BR & Coupling 
\\
\hline &&&&& \\[-15pt]
$S^L_0$ & 1 & $-{1\over 3}$ 
& $\begin{array}{c}e_L u \\ \nu_L d \end{array}$ 
& ${1\over 2}$ 
& $\begin{array}{c} g_L \\ -g_L \end{array}$ 
\\[6pt]
\hline &&&&& \\[-15pt]
$S^R_0$ & 2 & $-{1\over 3}$ 
& $e_R u $ 
& $1$ 
& $g_R$
\\[6pt] 
\hline &&&&& \\[-15pt]
$\tilde{S}^R_0$ & 3 & $-{4\over 3}$ 
& $e_R d $ 
& $1$ 
& $g_R$
\\[6pt]  
\hline &&&&& \\[-15pt]
$S^L_{1+}$ & 4 & $-{4\over 3}$ 
& $e_L d$  
& $1$ 
& $-\sqrt{2}\,g_L$  
\\[6pt] 
\hline &&&&& \\[-15pt]
$S^L_{10}$ & 5 & $-{1\over 3}$ 
& $\begin{array}{c}e_L u \\ \nu_L d \end{array}$  
& ${1\over 2}$ 
& $\begin{array}{c}  -g_L \\ -g_L \end{array}$  
\\[6pt] 
\hline &&&&& \\[-15pt]
$S^L_{{1\over 2}+}$ & 6 & $-{5\over 3}$ 
& $e_L \bar{u}$  
& $1$ 
& $g_L$
\\[6pt]   
\hline &&&&& \\[-15pt]
$S^R_{{1\over 2}+}$ & 7 & $-{5\over 3}$ 
& $e_R \bar{u}$  
& $1$ 
& $g_R$
\\[6pt]   
\hline &&&&& \\[-15pt]
$S^R_{{1\over 2}-}$ & 8 & $-{2\over 3}$ 
& $e_R \bar{d} $    
& $1$ 
& $ -g_R$  
\\[6pt] 
\hline &&&&& \\[-15pt]
$\tilde{S}^R_{{1\over 2}+}$ & 9 & $-{2\over 3}$ 
& $e_L \bar{d}$  
& $1$ 
& $g_L$  
\\[6pt] 
\hline 
\end{tabular}
\caption[.]{Scalar leptoquark charges, couplings and decay modes,
as used by {\tt ERATO-LQ}}
\label{choicess}
\ec
\end{table}

\begin{table}[ht]
\bc
\begin{tabular}{|c|c|c|c|c|c|}
\hline
LQ & {\tt ICHOICE} & Q & Decay Mode & BR & Coupling 
\\
\hline &&&&& \\[-12pt]
$ V^L_{{1\over 2}+}$ & 1 & $-{4\over 3}$ 
& $ e_L d $ 
& $1$ 
& $ g_L $ 
\\[6pt]
\hline &&&&& \\[-12pt]
$ V^R_{{1\over 2}+}$ & 2 & $-{4\over 3}$ 
& $e_R d $ 
& $1$ 
& $g_R$
\\[6pt] 
\hline &&&&& \\[-12pt]
$ V^R_{{1\over 2}-}$ & 3 & $-{1\over 3}$ 
& $e_R u $ 
& $1$ 
& $g_R$
\\[6pt]  
\hline &&&&& \\[-12pt]
$ \tilde{V}^R_{{1\over 2}+}$ & 4 & $-{1\over 3}$ 
& $e_L u $  
& $1$ 
& $g_L$  
\\[6pt] 
\hline &&&&& \\[-12pt]
$ V^L_0$ & 5 & $-{2\over 3}$ 
& $\begin{array}{c}e_L \bar{d} \\ \nu_L \bar{u} \end{array}$  
& ${1\over 2}$ 
& $\begin{array}{c} g_L\\ g_L \end{array}$  
\\[6pt] 
\hline &&&&& \\[-12pt]
$ V^R_0$ & 6 & $-{2\over 3}$ 
& $e_R \bar{d}$  
& $1$ 
& $g_R$
\\[6pt]   
\hline &&&&& \\[-12pt]
$ \tilde{V}^R_0$ & 7 & $-{5\over 3}$ 
& $e_R \bar{u}$  
& $1$ 
& $g_R$
\\[6pt]   
\hline &&&&& \\[-12pt]
$ V^L_{1+}$ & 8 & $-{5\over 3}$ 
& $e_L \bar{u} $    
& $1$ 
& $ \sqrt{2}\,g_L$
\\[6pt]  
\hline &&&&& \\[-12pt]
$ V^L_{10}$ & 9 & $-{2\over 3}$ 
& $\begin{array}{c}e_L \bar{d} \\ \nu_L \bar{u} \end{array}$    
& ${1\over 2}$ 
& $\begin{array}{c} - g_L\\ g_L \end{array}$   
\\ 
\hline
\end{tabular}
\caption[.]{Vector leptoquark charges, couplings and decay modes,
as used by {\tt ERATO-LQ}}
\label{choicesv}
\ec
\end{table}

\vspace*{3cm}

\end{document}